\documentclass[prl,aps,twocolumn,superscriptaddress]{revtex4-2}            
\usepackage[utf8]{inputenc}
\usepackage[T1]{fontenc}                
\usepackage{amsmath, amsfonts, stmaryrd,dsfont,graphicx,hyperref,bm,mathtools,verbatim,color}
\usepackage[caption=false]{subfig}
\usepackage[export]{adjustbox}
\usepackage[capitalise]{cleveref}
\def\mb{\mathbf}
\def\a{\alpha}
\def\b{\beta}

\def\r{\rho}

\def\6{\partial} 
\def\d{\delta}

\def\g{\gamma}

\def\p{\pi}

\newcommand{\ea}[1]{\begin{align}#1\end{align}}
\newcommand{\eq}[1]{\begin{equation}\begin{split}#1\end{split}\end{equation}}

\def\l{\lambda}

\begin{document}
	%\title{Entanglement entropy between antiferromangetic sublattices}
	\title{Intersublattice entanglement entropy as an extensive property in antiferromagnets}
	\author{Dion M. F. Hartmann}\email[E-mail adress: ]{d.m.f.hartmann@uu.nl}\affiliation{Institute for Theoretical Physics, Utrecht University, Leuvenlaan 4, NL-3584 CE Utrecht, The Netherlands}
	\author{Jurriaan J. Wouters}\affiliation{Institute for Theoretical Physics, Utrecht University, Leuvenlaan 4, NL-3584 CE Utrecht, The Netherlands}
	\author{Dirk Schuricht}\affiliation{Institute for Theoretical Physics, Utrecht University, Leuvenlaan 4, NL-3584 CE Utrecht, The Netherlands}
	\author{Rembert A. Duine}\affiliation{Institute for Theoretical Physics, Utrecht University, Leuvenlaan 4, NL-3584 CE Utrecht, The Netherlands}\affiliation{Center for Quantum Spintronics, Department of Physics, Norwegian University of Science and Technology, NO-7491 Trondheim, Norway}\affiliation{Department of Applied Physics, Eindhoven University of Technology, P.O. Box 513, 5600 MB Eindhoven, The Netherlands}
	\author{Akashdeep Kamra}\affiliation{Center for Quantum Spintronics, Department of Physics, Norwegian University of Science and Technology, NO-7491 Trondheim, Norway}
	\begin{abstract}
		Recent advancements in our understanding of ordered magnets call for a quantification of their entanglement content on an equal footing with classical thermodynamic quantities, such as the total magnetic moment. We evaluate the entanglement entropy (EE) between the two sublattices of a bipartite ordered antiferromagnet finding it to scale with volume. Thus, the EE density becomes an intensive property and is evaluated to be a universal dimensionality-dependent constant when exchange is the dominant interaction. Our analytic results are validated against the DMRG-based analysis of a one-dimensional (1D) system, finding good agreement. Further, our evaluated EE per bond provides a useful shortcut towards obtaining the central-cut EE in 1D, and the area law in higher-dimensional magnets. 
		%Our work should facilitate EE-based analysis of antiferromagnetic ground states and exploitation of industrial magnets {\color{red} in generating entanglement for quantum protocols}.
	\end{abstract}
\pacs{03.65.Ud,03.67.Bg,75.10.Jm,75.30.Ds,75.50.Ee}
\maketitle	
\textit{Introduction}. --- Antiferromagnets (AFMs) and their different phases pervade condensed matter physics. They underlie research fields, such as spin fluctuations mechanism of high-$T_\textrm{c}$ superconductivity \cite{mor00,mor03,Lee2006} and quantum spin liquids \cite{ito10,han12,Balents2010,Savary2016,Castelnovo2008}, as well as applications, such as exchange-biasing in magnetic read heads~\cite{Zhang2016,Manna2014,Nogues1999}. There exists a sharp contrast between the two widespread approaches towards understanding AFMs. In the first ``quantum'' approach~\cite{Balents2010,Savary2016,Castelnovo2008}, determining the ground state wavefunction for various model AFMs is a major goal. The true ground state, comprising distant entangled spins, is often not known and is complicated. Various numerical methods are employed in approximating the ground state and the excitations. The investigated system size is often limited by computational power. In the second ``semiclassical'' approach~\cite{Gilbert2004,Baltz2018,Gomonay2014,Gomonay2018,Kamra2018B,Jungwirth2016}, a mean-field approximation is made and spatially resolved spin densities or magnetizations become classical fields within the Landau-Lifshitz description~\cite{Holstein1940,Akhiezer1968}. A N\'eel ordered ground state is assumed and yields results consistent with many experiments~\cite{Lebrun2018,Baltz2018,Gomonay2018,Jungwirth2016}. Macroscopic averaging is one of the reasons why nonlocal spin correlations and entanglement, fundamental in the quantum approach, appear to not affect several experiments consistent with the semiclassical approach.
	
Entanglement is an important resource in quantum information and computing protocols~\cite{Ralph1999,Ou1992,Furrer2012}. Two subsystems are said to be entangled if the wavefunction describing the total system cannot be factored into a product of two wavefunctions, one for each subsystem. Further, in the quantum approach discussed above, entanglement offers a powerful metric for characterizing ground state and excitations, mapping complicated wavefunctions existing in very high-dimensional spaces to a scalar quantity~\cite{Savary2016,Nishioka2018,Islam2015,ami08}. In a widely employed technique, a 3-dimensional AFM is partitioned via a closed surface and the entanglement entropy (EE) between the two partitions is evaluated. The EE then bears a contribution proportional to the partition surface area, known as the area law \cite{cal09,eis10,Savary2016,ami08}. An additional, and sometimes universal, offset in the area law probes and characterizes topology and long-range entanglement in ground and excited states of the AFM~\cite{Kitaev2006,ami08}.
	
Recent works relate magnons with squeezed states studied in quantum optics~\cite{Kamra2016A,kam19,Zou2020,Kamra2020} to demonstrate a nonzero entanglement in magnets~\cite{kam19,Zou2020,Yuan2020,Kamra2020}, even in the mean-field approximation employed in the magnon-based semiclassical theory. This calls for a systematic quantification of EE as a quantum property describing such ordered magnets. We note two key motivators for this. First, the EE offers a simple scalar metric from which the proximity of a numerically evaluated ground state wavefunction of an ordered state can be measured. This facilitates analysis and approximations for quantum ground states. Second, the recent breakthroughs in robust experimental control of AFMs~\cite{Baltz2018,Jungwirth2016,Lebrun2018} obeying the semiclassical approach outlined above pave the way for using them as a resource or battery for entanglement~\cite{ami08,Zou2020,Awschalom2021}. Such efforts benefit from adding EE to the (quantum) thermodynamic description of magnets. 
	
\begin{figure}
	\includegraphics[width=\linewidth]{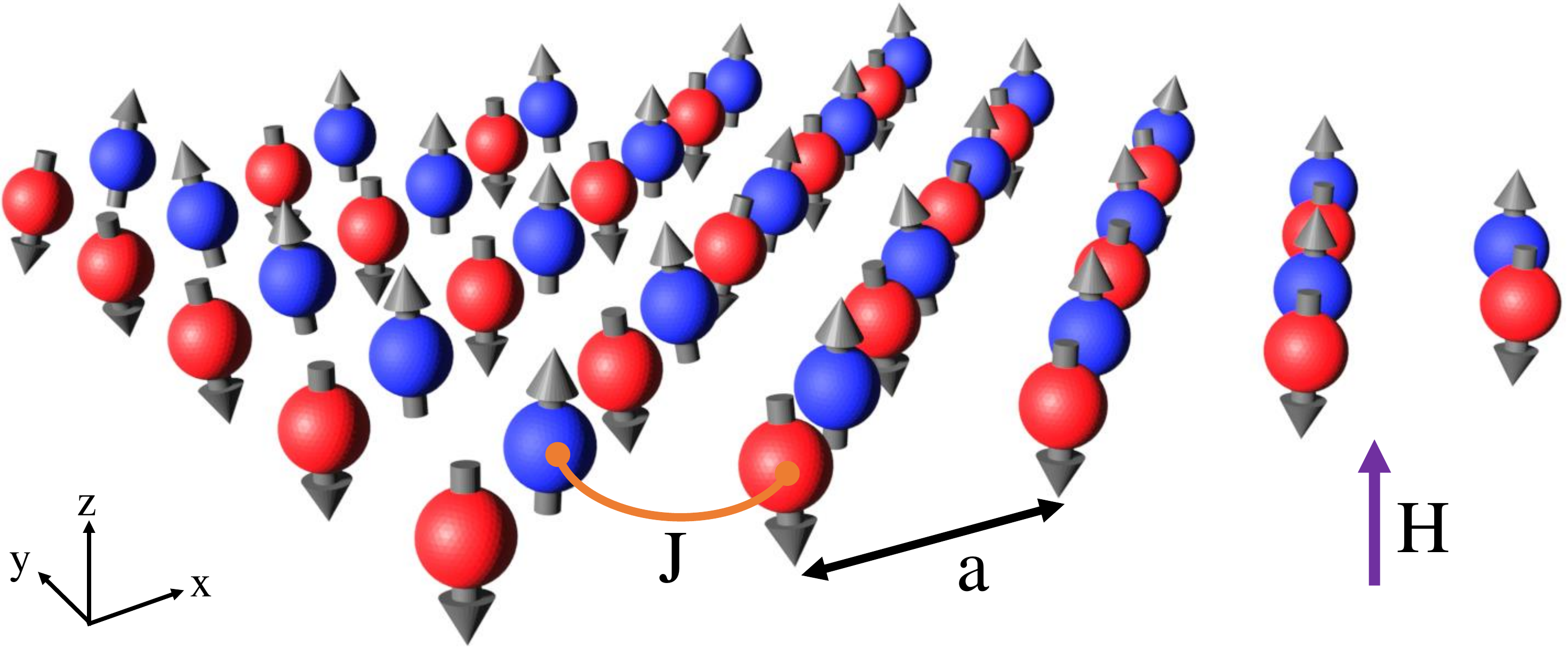}
	\caption{
		\label{fig:sys}
		(Color online) 
		The studied system for $d=2$, $N=5$ close to a Néel state. Sublattice A (blue) and B (red) both have an equal amount of spins. The exchange interaction (orange) is restricted to nearest neighbors and is identical for each pair of neighbors. The external magnetic field (purple) is along the $z$ direction.
	}
\end{figure}

In this paper, partitioning the AFM into two sublattices (\cref{fig:sys}), we establish the EE (density) as an extensive (intensive) quantum property characterizing ordered AFMs. Working within the mean-field approximation and magnon picture, we analytically evaluate the EE in the ground state, finding it to scale with the system size in the thermodynamic limit. Contradicting a preliminary expectation suggesting an increase in EE with exchange interaction strength~\cite{kam19,Kamra2020}, the EE density is found to be a universal constant depending only on system dimensionality. This universality could offer useful benchmarking in analyzing quantum ground states. Examining its dependence on an applied magnetic field, we find that EE remains unchanged on approaching the spin-flop transition, where various classical response functions diverge~\cite{Akhiezer1968,Johansen2017}. 
%This reaffirms the unique role of EE in capturing only quantum fluctuations and correlations. 
Our analytic results are found to agree well with a density matrix renormalization group (DMRG) analysis of a 1D AFM. Further, our evaluated EE per bond provides an analytic shortcut to evaluating EE for the widely employed system partitioning into two spatially separated regions~\cite{ami08}.

\textit{Model}. --- 
We consider a $d$-dimensional uniaxial AFM in an external magnetic field along the $z$-axis with $N$ spins in each direction on sublattice $A$ ($B$) pointing along the $(-)z$-axis described by the Hamiltonian
\eq{
	\mathcal{H}=&\frac{J}{\hbar^2}\sum_{i,\bm{\d}}\bm{S}_A(\mb{r}_i)\cdot\bm{S}_B(\mb{r}_i+\bm{\d})
	\\
	&-\sum_{\substack{\a\in\{A,B\}\\i}}\left(\frac{K}{\hbar^2}\left(S^{(z)}_\a(\mb{r}_{i_\a})\right)^2+\gamma H S^{(z)}_\a(\mb{r}_{i_\a})\right)
	,
}
with $J$ the exchange coupling, $K$ the anisotropy energy, $\bm{\d}$ the vectors to nearest neighbors, $\g<0$ the gyromagnetic ratio and $H$ the external magnetic field. Our final expression for EE does not depend on the lattice or spin $S$ considered, as long as $\sum_{\bm\d}\bm{\d}=\mb{0}$. Thus, for concreteness and without loss of generality, we consider a square lattice. \cref{fig:sys} depicts a 2-dimensional system with $N=5$ spins per sublattice per dimension. Assuming the magnetic field is below the spin-flop transition [i.e. $|\gamma|\hbar H< 2S\sqrt{K(J+K)}$], we apply a Holstein-Primakoff transformation to express the Hamiltionan in local bosonic operators $a_i$ ($b_j$) which annihilate a spin flip on the $i$-th ($j$-th) site of the $A$ ($B$) sublattice and satisfy the canonical bosonic commutation relations \cite{hol40}. We assume periodic boundary conditions. After a Fourier transform we obtain up to second order in these ladder operators
\ea{\hspace{-2mm}
	\mathcal{H}=\sum_\mb{k}A_+ a_\mb{k}^\dagger a_\mb{k}+A_-b_\mb{k}^\dagger b_\mb{k}+C_\mb{k}a_\mb{k}b_{-\mb{k}}+C^\star_\mb{k}b_{-\mb{k}}^\dagger a_\mb{k}^\dagger;
	\\
	\hspace{-2mm}
	\label{eq:ACk}
	A_\pm=JSc+2KS\pm\g \hbar H; \qquad C_\mb{k}=JS\sum_{\bm{\d}}e^{i\mb{k}\cdot\bm{\d}}.
}
Here, $c$ is the coordination number and $S$ the total spin per site. The sum over $\mb{k}$ runs over the Brillouin zone, i.e. $\mb{k}\cdot\hat{\mb{e}}_i\in\{-\p/a,...,\p(N-2)/L\}$, with $L=aN$ and $a$ the lattice spacing. It is known \cite{kam17} that the eigenstate depends only on the sum $A_++A_-\equiv 2A$, hence the magnetic field does not affect the entanglement entropy within the Néel state approximation. 

We diagonalize the Hamiltonian by applying a Bogoliubov transformation $\mathcal{H}=\sum_\mb{k}E_\mb{k}(\a_\mb{k}^\dagger \a_\mb{k}+\b_\mb{k}^\dagger \b_\mb{k})$ to obtain the eigen energy
%\eq{
%	\mathcal{H}=\sum_\mb{k}E_\mb{k}\left(\a_\mb{k}^\dagger \a_\mb{k}+\b_\mb{k}^\dagger \b_\mb{k}\right),
%} 
$E_\mb{k}=-|\g|\hbar H+E_\mb{k}^0$, with $E_\mb{k}^0=\sqrt{A^2-C_\mb{k}^2}$ and $\a_\mb{k}$ and $\b_\mb{k}$ are squeezed sublattice magnons:
\eq{
	\label{eq:abk}
	\a_\mb{k}=u_\mb{k}a_\mb{k}+v_\mb{k}b_{-\mb{k}}^\dagger,\qquad 
	&\b_\mb{k}=u_\mb{k}b_\mb{k}+v_\mb{k}a_{-\mb{k}}^\dagger,
	\\
	u_\mb{k}=\sqrt{\frac{A+E_\mb{k}^0}{2E_\mb{k}^0}},\qquad 
	&v_\mb{k}=\sqrt{\frac{A-E_\mb{k}^0}{2E_\mb{k}^0}}.
}
Thus the ground state is squeezed \cite{ger05}: Whereas $a$ and $b$ operate only on one sublattice, $\a$ and $\b$ operate on both. So we expect a finite entanglement in the ground state.

In Fock notation we write a state in terms of $\otimes_\mb{k}|N_{b_\mb{k}},N_{a_\mb{k}}\rangle_{sub,\mb{k}}$ in the sublattice basis, or in terms of $\otimes_\mb{k}|N_{\b_\mb{k}},N_{\a_\mb{k}}\rangle_{sq,\mb{k}}$ in the squeezed basis. 
The product over $\mb{k}$ ranges over all $N^d$ allowed wave modes.
The ground state is the squeezed vacuum $\otimes_\mb{k}|0,0\rangle_{sq,\mb{k}}$. 
We use the two-mode squeezing operator $S(r_\mb{k})=\exp r_\mb{k}\left(a_\mb{k} b_\mb{k}-a_\mb{k}^\dagger b_\mb{k}^\dagger\right)$, with $u_\mb{k}=\cosh r_\mb{k}$ and $v_\mb{k}=\sinh r_\mb{k}$, and exploit the Baker-Hausdorff lemma \cite{sak95}
%\eq{
%	S(r_\mb{k})a_\mb{k}S(r_\mb{k})^\dagger=\a_\mb{k};\qquad
%	S(r_\mb{k})b_\mb{k}S(r_\mb{k})^\dagger=\b_\mb{k},
%}
to express the ground state in terms of the sublattice basis 
\eq{\label{eq:GSsq}
	|G\rangle=&\otimes_\mb{k}|0,0\rangle_{sq,\mb{k}}=\prod_\mb{k}S(r_\mb{k})\otimes_\mb{k}|0,0\rangle_{sub,\mb{k}}\\
	=&\otimes_\mb{k}\frac{1}{\cosh r_\mb{k}}\sum_{l=0}^\infty(-\tanh r_\mb{k})^l|l,l\rangle_{sub,\mb{k}}.
}
So, for each mode $\mb{k}$ we have a sum over $l$ ranging over all occupation numbers of this mode.
%\newline

\textit{Entanglement Entropy}. --- 
%The Baker-Hausdorff lemma yields
%\eq{
%	S(r_\mb{k})a_\mb{k}S(r_\mb{k})^\dagger=\a_\mb{k};\qquad
%	S(r_\mb{k})b_\mb{k}S(r_\mb{k})^\dagger=\b_\mb{k}
%}
Using the Schmidt decomposition we derive the reduced density matrix 
\eq{
	\r_A=\textrm{Tr}_B\r=&\sum_\mb{n} \prescript{}{B}{\langle}\mb{n}|G\rangle\langle G|\mb{n}\rangle_B
	=\sum_\mb{l} \prescript{}{B}{\langle}\mb{l}|G\rangle\langle G|\mb{l}\rangle_B\\
	=&\otimes_\mb{k}\left(\sum_l\frac{\tanh^{2l}r_\mb{k}}{\cosh^2r_\mb{k}}|l\rangle_{A,\mb{k}}\prescript{}{A,\mb{k}}{\langle}l|\right),
}
where $\mb{n}=(n_1,n_2,...,n_{N^d})$ is a vector of $N^d$ integers used to determine a pure Fock state $|\mb{n}\rangle_B$ on the $B$ sublattice in the position basis, i.e. each integer $n_i$ gives the occupancy of the $i$-th position on the $B$ sublattice. Similarly, $\mb{l}$ is a vector used to determine a pure Fock state on the $B$ sublattice in the momentum basis. 

\begin{figure}[t]
	\centering
	\subfloat[\label{fig:Sofk1D}]{ \includegraphics[width=0.47 \linewidth,valign=b]{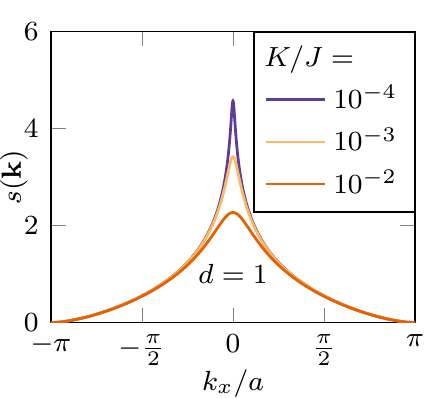}}
	\hfill
	\subfloat[\label{fig:Sofk2D}]{ \includegraphics[width=0.5 \linewidth,valign=b]{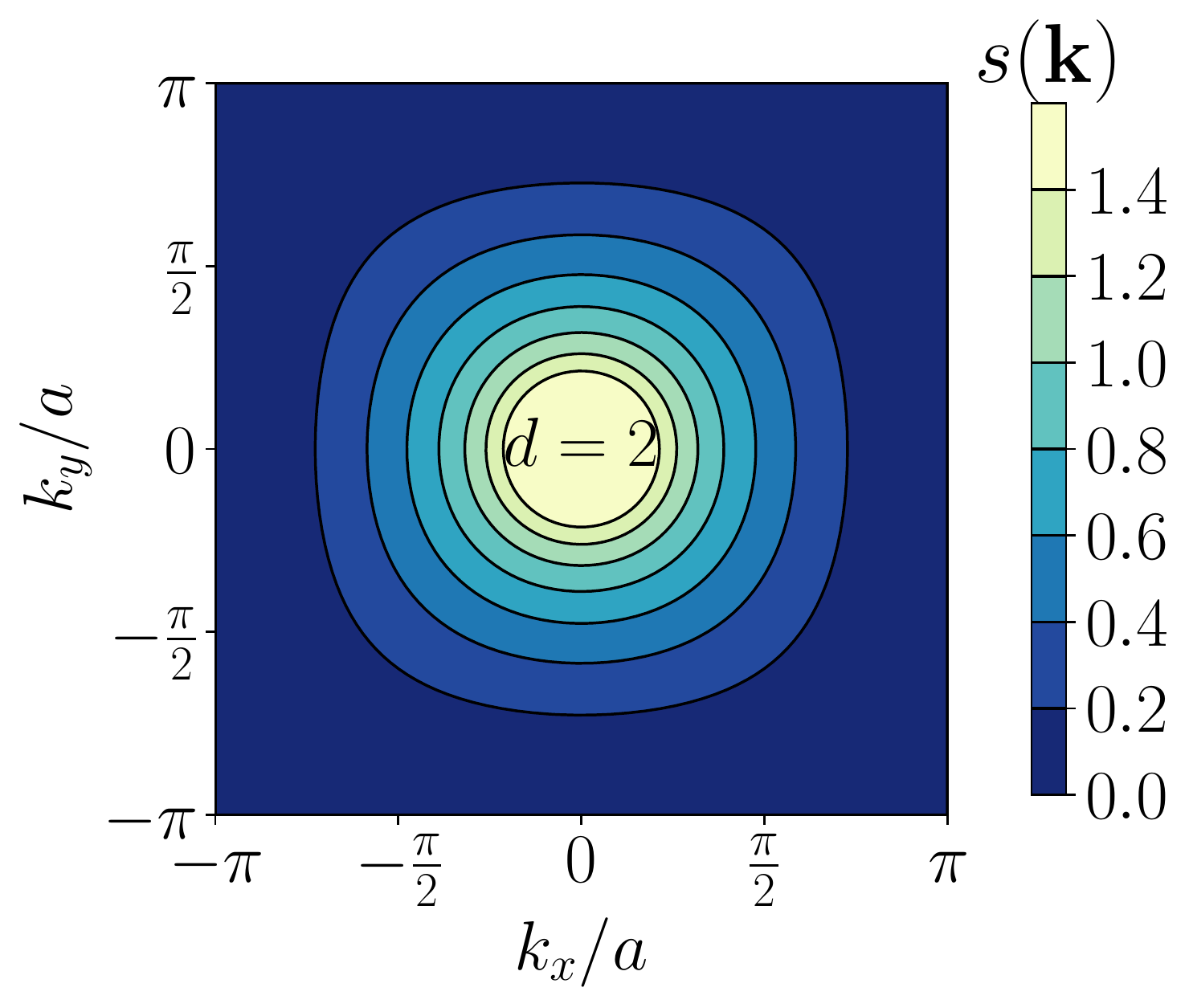}}
	\caption{(Color online) The entanglement entropy per wavenumber $s(\mb{k})$ in $d=1$ (a) and $d=2$, (b) with $K/J=10^{-4}$. Note that near the Brillouin zone boundaries, the EE vanishes. Whereas in the low wavenumber regime, the EE depends approximately only on the radius of the wavevector. Thus, a smaller contribution from the high-$k$  modes allows a low-$k$ continuum approximation to analytically obtain the total EE. \label{fig:sok}}
\end{figure}

\begin{figure}[!h]
	\centering
	\subfloat[\label{fig:SvsS_N}]{ \includegraphics[height=0.4 \linewidth,valign=t]{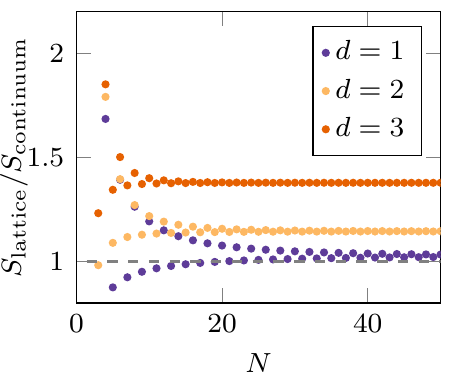}}
	\hfill
	\subfloat[\label{fig:SvsS_K}]{ \includegraphics[height=0.4	 \linewidth,valign=t]{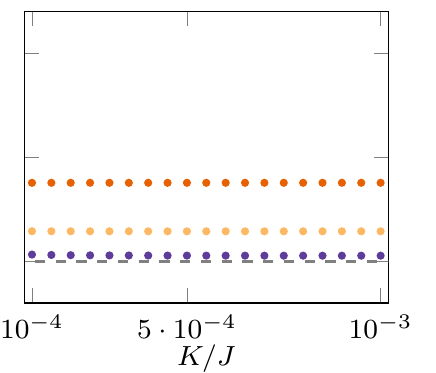}}
	\caption{(Color online) The lattice entanglement entropy [\cref{eq:Sec}] divided by the approximated continuum result [\cref{eq:Sap}], plotted as a function of the system size (a) and the ratio $K/J$ for $N=50$ (b) for $d=1$ (blue), $d=2$ (yellow) and $d=3$ (orange). \label{fig:svs}}
\end{figure}

The eigenvalues of the reduced density matrix follow immediately
\eq{
	\r_A|\mb{n}\rangle_A=\l_\mb{n}|\mb{n}\rangle_A;\qquad
	\l_\mb{n}=\prod_{i=1}^{N^2}\frac{\tanh^{2n_i}r_{k_i}}{\cosh^{2}r_{k_i}}.
}
This yields an analytic expression for the EE 
\eq{
	S_\textrm{EE}&=
	-\sum_\mb{n}\l_\mb{n}\log\l_\mb{n}\\
	&=
	\sum_\mb{k}2\left(\log\cosh r_\mb{k}-\sinh^2r_\mb{k}\log\tanh r_\mb{k}\right).\label{eq:Sex}
	%	\\
	%	&=
	%	\sum_\mb{k}u_\mb{k}^2\log u_\mb{k}^2-v_\mb{k}^2\log v_\mb{k}^2.
}
%\begin{figure}
%	\includegraphics[width=\linewidth]{Sofk_old}
%	\caption{
%		\label{fig:sok}
%		The entanglement entropy density (per wavenumber) $s(\mb{k})$ in $d=1$ (a) and $d=2$ (b). Note that near the Brillouin zone boundaries, the EE vanishes. Whereas in the low wavenumber regime, the EE approximately depends only on the radius of the wavevector.
%	}
%\end{figure}
Note that in the sum over $\mb{k}$ all components $k_i$ range from $-\p/a$ to $\p(N-2)/L$ in $N$ steps of $2\p/L$. The squeezing parameter $r_{\bm{k}}$ (and thereby the EE) does not depend on the applied external magnetic field when the magnetic field is below the spin-flop strength. Furthermore, there also is no $S$ dependence. For brevity we denote the term in the sum as $s(\mb{k})$ to investigate scaling behavior in the large $N$ limit
\eq{
	\label{eq:Sec}
	S_\textrm{EE}=\sum_\mb{k}s(\mb{k})\approx L^d\int \frac{d^d\mb{k}}{(2\p)^d}s(\mb{k}).
}
From \cref{eq:ACk} we see that the only $\mb{k}$ dependence lies in $C_\mb{k}$. As \cref{fig:sok} demonstrates for the one- and two-dimensional system, for $\mb{k}$ approaching the Brillouin zone boundary the contribution to the EE vanishes since $r_\mb{k}\rightarrow 0$. 
%This suggests that the specific structure of the lattice does not significantly affect the EE. To further investigate this claim,
Furthermore, note that for small $\mb{k}$ the EE density depends mostly on the norm of $\mb{k}$. This warrants us to consider the small $\mb{k}$ limit and expand
\eq{
	\label{eq:ckd}
	\frac{|C_\mb{k}|^2}{A^2}\approx\frac{1}{(1+\frac{2K}{cJ})^2}\left(1-\frac{a^2 k^2}{4}\right),
}
with $k^2=\sum_{i=1}^d k_i^2$ and for the square lattice $c=2^d$, $\bm{\d}\cdot\hat{\mb{e}}_i=\pm a/2$ and $C_\mb{k}$ is real. This is the only point in our derivation where details of the lattice structure enter. For instance, for a honeycomb lattice we would have $c=3$.
% with the same nearest neighbor spacing $a/\sqrt{2}$:
%\eq{
%	\label{eq:ckdhex}
%	\frac{|C_\mb{k}|^2}{A^2}\approx\frac{1}{(1+\frac{2}{3}K/J)^2}\left(1-\frac{a^2 k^2}{4}\right).
%}
This demonstrates that for $K\ll J$ our result is universal with respect to the lattice structure.
Transforming the integral to spherical coordinates, leaving only the integral over the radial component $k$, we obtain
\eq{
	\label{eq:Sap}
	S_\textrm{EE}\approx 2^{\frac{d}{2}+1}N^d\frac{(1+\frac{K}{2^{d-1}J})^{d/2}}{\p^{d/2}\Gamma (\frac{d}{2})}I\left(d,\frac{K}{J}\right),
}
where $\Gamma$ is the Euler gamma function and an integral representation for $I$ is given in the supplemental material \cite{SM} where we derive analytically for $K\ll J$ that $I(d,K/J)\sim d^{-2}$.
%with $I(d,10^{-4})\approx 0.94 d^{-2}$ and $\Gamma$ the Euler Gamma function. Details on $I$ are given in the supplemental materials \cite{SM}. 
In \cref{fig:svs} the ratio between the lattice result from \cref{eq:Sec} and the continuum limit from \cref{eq:Sap} is plotted versus $N$ (a) and the ratio $K/J$ (b). Note that the finite size effects quickly vanish as $N$ increases and that the EE is independent of the ratio $K/J$ in the regime $K/J\ll 1$. The error of the continuum result increases with $d$ as it is caused by the increasing inaccuracy of the small $\mb{k}$ approximation.
%\ea{
%	I(d,c)=&\int_0^{\p/2\sqrt{2\left(1+2^{1-d}c\right)}} dx x^{d-1} \\
%	\nonumber
%	& g_k^+(d,c)\log g_k^+(d,c)-g_k^-(d,c)\log g_k^-(d,c);
%	\\
%	g_k^\pm(d,c)=&\frac{1}{2}\left[\left(1-\left(\frac{1}{1+2^{1-d}c}-k^2\right)^2\right)^{-1/2} \pm1\right].
%}
%\begin{figure}[t!]
%	\includegraphics[width=\linewidth]{SvsS_old}
%	\caption{
%		\label{fig:svs}
%		The exact entanglement entropy divided by the result from our analytical approach, plotted as a function of the system size (a) and the ratio $K/J$ (b) for $d=1$ (blue), $d=2$ (orange) and $d=3$ (purple). [[figure (b) over a larger regime? Say something about the inaccuracy?]]
%	}
%\end{figure}

\begin{figure}[!h]
	\includegraphics[width=\linewidth]{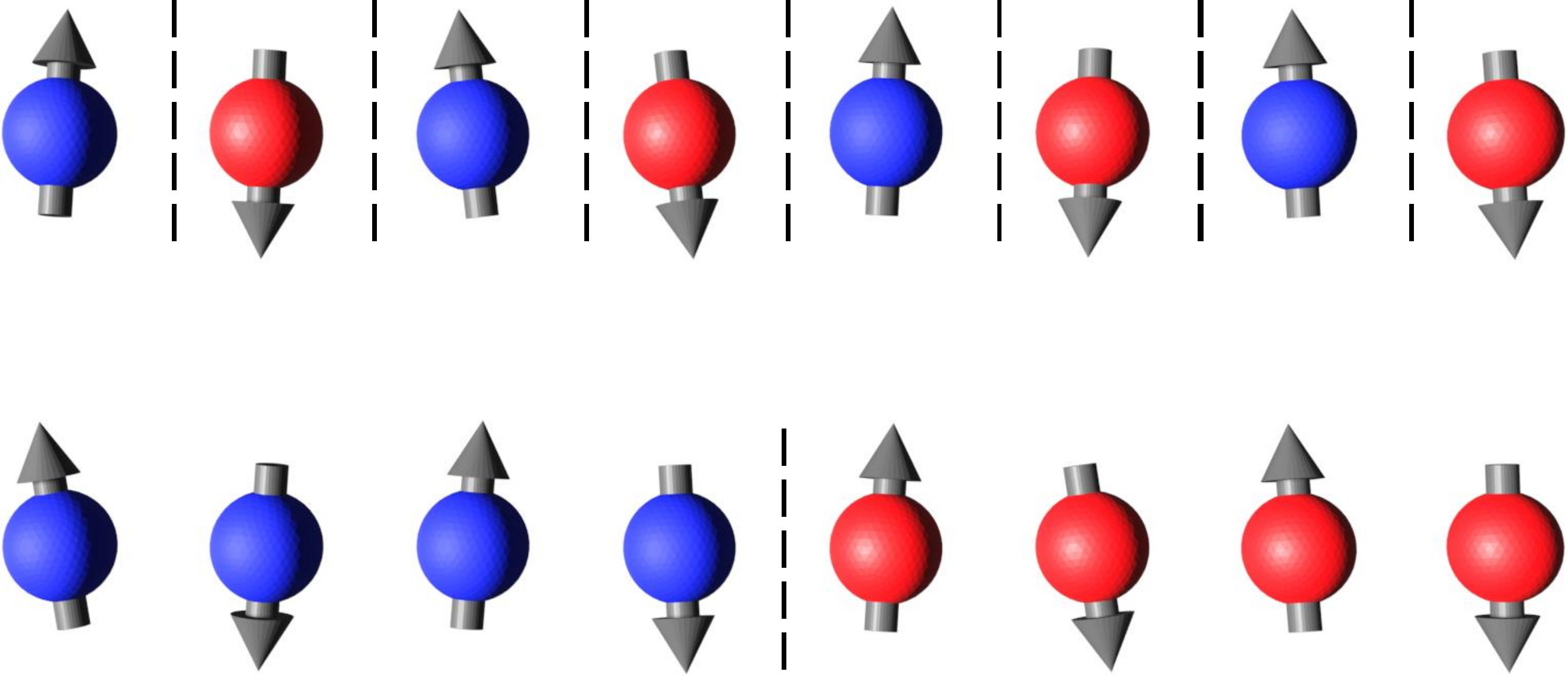}
	\caption{
		\label{fig:CCEEC}
		(Color online) For the intersublattice entanglement (top) we divide our system into two parts, red and blue, by alternating sites. For the central cut entanglement(CCEE) (bottom) we split the system through the middle. As the DMRG results suggest, the CCEE is a good approximation of the intersublattice entanglement entropy per bond.
	}
\end{figure}

\textit{DMRG for a 1D chain}. --- 
The DMRG method \cite{whi93} is a versatile variational numerical tool to find the low energies and corresponding eigenstates of a strongly correlated 1D system in polynomial (in system size) time. 
%First proposed by Steven White in 1992 \cite{whi93}, it has positioned itself as the go-to numerical tool for 1D lattice systems. It was later realized that White's method has a very natural representation 
The DMRG method is formulated
in terms of matrix product states (MPS) \cite{sch11}, making the EE straightforwardly accessible. 
%The MPS allow for a straightforward interpretation to the non-trivial entanglement entropy of the system. For a review see Ref.~\cite{sch11}. 

The DMRG algorithm approaches the lowest energy state by optimizing the MPS locally, alternating over all sites. It retains only the $D$ most relevant states, selecting them based on the highest singular value/weight $s_i$. Normalization of the state requires $\sum_i s_i^2=1$.  The bond dimension $D<D_{\rm max}$ is set such that the weight of the discarded state is $\sum_{j>D}s_j^2<10^{-5}$. From the von Neumann entropy $S_{\rm EE}=-2\sum_{i=1}^Ds_i^2\log(s_i)$ we see that the minimal required bond dimension is related to the EE. In contrast with the two-sublattice partitioning discussed above, the DMRG utilizes the bipartition displayed in the bottom of Fig.~\ref{fig:CCEEC}. The area law tells us that, for gapped systems, the entanglement and thus the bond dimension is chosen independent of system length \cite{cal04,cal09,eis10}. This showcases the great advantage of DMRG, requiring only linear (in $N$) memory allocation and polynomial computation time. The numerical results presented here are obtained with the TeNPy library\cite{hau18}.

\begin{figure}
	\includegraphics[width=\linewidth]{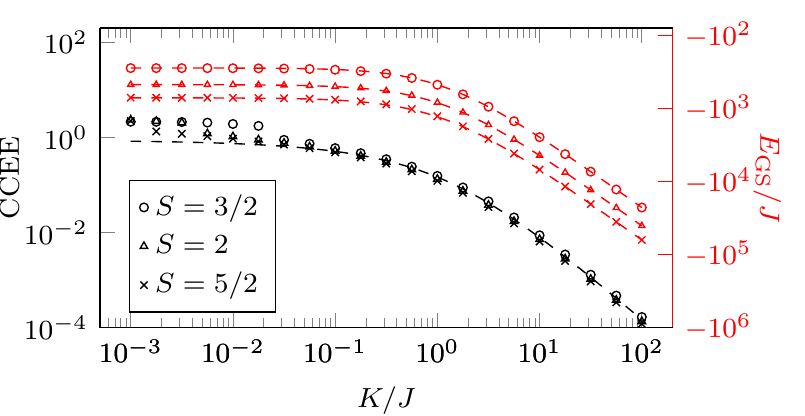}
	\caption{
		\label{fig:EE_AND_E_ground_state}
		(Color online) 
		Comparison of approximate analytical ground state \eqref{eq:GSsq} with the full DMRG ground state for $N=50$. The marks denote the numerical results for spin-3/2 (circles), spin-2 (triangles) and spin-5/2 (crosses) with the analytical result as a dashed line. The black results compare the central-cut entanglement entropy (CCEE) for DMRG to the analytical spatial entanglement entropy density $S_\textrm{EE}/2N$. The red results compare the analytical with the numerical ground state energy.
	}
\end{figure}

The area law in 1D dictates a linear relation of the EE to the length of the cut: $S_{\rm EE}=S_0 L_{\rm cut}$. We recognize for the central-cut bipartition (bottom Fig.~\ref{fig:CCEEC}) $L_{\rm cut}=1$ and for intersublattice biparition (top Fig.~\ref{fig:CCEEC}) $L_{\rm cut}=2N$. This allows us to compare the analytic intersublattice EE with the numerical CCEE. The black data in Fig.~\ref{fig:EE_AND_E_ground_state} confirm that the numerical CCEE matches the intersublattice EE density (no spin dependence) very well for spin $S=3/2,2,5/2$ and a large range of anisotropy ($K/J$), moreover confirming numerically the independence of EE with respect to $S$.
We attribute the deviation in the CCEE at small $K/J<5\cdot 10^{-2}$ to the existence of low-energy modes. Specifically, for half-integer spin $S$ the model at $K=H=0$ is gapless by virtue of the Lieb-Schultz-Mattis theorem \cite{lie61,tas20}. On the other hand, for $S=2$ a phase transition was identified at \cite{kja13} $K/J\approx 0.0046$. In both cases, low-energy states are present, leading to a logarithmic dependence of the CCEE on the system size \cite{cal04,cal09}.

Besides the good agreement in EE, Fig.~\ref{fig:EE_AND_E_ground_state} shows in red that the analytical and numerical ground state energies match, giving yet another hint that the squeezed state provides a good representation of the low energy physical behavior.

\textit{Discussion}. --- Some key features of and a comparison between our analytic and DMRG methods should be noted. In the former approach, the two-sublattice partitioning allowed us to express the total EE as a sum over $\mb{k}$ [\cref{eq:Sec}]. This further allowed demonstrating its scaling with system volume and obtaining analytic results [\cref{eq:Sap}] via the continuum approximation. As the EE becomes small for $\mb{k}$ close to the Brillouin zone boundary, the dependence of EE on the microscopic lattice is expected to be weak, as motivated in \cref{fig:Sofk2D}. Thus we expect the result \cref{eq:Sec} to be valid for bipartite lattices in general, and in this sense to be universal. 

Crucial to these simplifications has been our unconventional choice of the partitioning which admits translation invariance, see \cref{fig:sys}. Due to the equivalence between all nearest neighbour exchange ``bonds'', the total EE per bond becomes a well-defined quantity. Thus, the area law of EE obtained with more commonly employed partitions dividing the AFM into two parts~\cite{Savary2016,ami08}, is understood as the EE per bond times the number of bonds that the partition intersects. This is illustrated in \cref{fig:CCEEC}. In this manner, our finding of EE being an extensive property is consistent with the area law and our choice of the partition~\cite{cal09,ami08}. 

Further, we find that EE does not depend on the applied magnetic field and remains constant [see \cref{eq:abk}], as the system approaches the spin-flop transition from below. At this value of applied magnetic field, one of the magnon modes becomes gapless resulting in a divergence in various response function~\cite{Akhiezer1968,Johansen2017,Lebrun2018}, such as the high-frequency susceptibility. Nevertheless, the EE remains well behaved and unperturbed as it is a property of the ground state wavefunction, which remains unaltered on approaching this transition from below.

Our DMRG based results show a good agreement with respect to the energy and EE of the analytic squeezed ground state as displayed in \cref{fig:EE_AND_E_ground_state}. While the agreement between the energies is excellent, a small deviation in the EE demonstrates that it is more sensitive when comparing the quantum ground states. Furthermore, a good agreement between our analytic EE per bond and the central-cut EE for long chains suggests a shortcut in evaluating the latter, disregarding the finite size effects. Hence, our two-sublattice partitioning may be useful beyond the ordered ground states considered here.

\textit{Summary}. --- We have investigated the entanglement entropy in the ground state of an ordered antiferromagnet using a two-sublattice partitioning. The translational invariance associated with the latter enabled us to obtain analytic results, consistent with numerics, providing insights and shortcuts in characterizing the entanglement content. Our finding of a universal behavior of the entropy helps benchmark numerically evaluated quantum ground states and guide the development of ordered antiferromagnets for useful quantum information protocols.

\begin{acknowledgments}
	R.D. and D.S. are members of the D-ITP consortium, a program of the Dutch Organization for Scientific Research (NWO) that is funded by the Dutch Ministry of Education, Culture and Science (OCW). R.D. and D.H. have received funding from the European Research Council (ERC) under the
	European Unions Horizon 2020 research and innovation programme (Grant agreement No. 725509).
	This work is funded by the European Research Council (ERC) and the Research Council of Norway through its Centers of Excellence funding scheme, project 262633, ``QuSpin''. This work is part of the research programme of the Foundation for Fundamental Research on Matter (FOM), which is part of the Netherlands Organization for Scientific Research (NWO).	
\end{acknowledgments}

\bibliography{References_MagnonEE}
\end{document}